\documentclass[envcountsame]{llncs}

\usepackage[utf8]{inputenc}
\usepackage[T1]{fontenc}
\usepackage{enumerate}
\usepackage{graphicx}
\usepackage{amsfonts}
\usepackage{amssymb}
\usepackage{amsmath}
\usepackage{wasysym}
\usepackage{hyperref}
\usepackage{mathtools}
\usepackage{bibentry}
\usepackage{comment}
\usepackage{enumitem}
\usepackage{todonotes}
\usepackage{hyperref}
\usepackage{lineno}

\usepackage{apxproof}

\newtheorem{observation}[theorem]{Observation}{\bfseries}{\itshape}

\def\computationproblem#1#2#3{
  \begin{center}
  \begin{tabular}{rp{0.8\textwidth}}
  {\sc Problem:\enspace}&#1\\
  {\sc Input:\enspace}&#2\\
  {\sc Question:\enspace}&#3\\
  \end{tabular}
  \end{center}
}

\let\proof\relax
\usepackage{amsthm}

\title{Computational Complexity of Covering Disconnected Multigraphs\thanks{The conference version of this paper appeared in the proceedings of FCT 2021~\cite{BFJKS21FCT}.}}

\author{Jan Bok\inst{1}\orcidID{0000-0002-7973-1361} \and Jiří Fiala\inst{2}\orcidID{0000-0002-8108-567X} \and Nikola Jedličková\inst{2}\orcidID{0000-0001-9518-6386} \and Jan Kratochvíl\inst{2}\orcidID{0000-0002-2620-6133} \and Michaela Seifrtová\inst{2}\orcidID{0000-0003-0050-480X}}

\institute{
University of Clermont Auvergne, LIMOS, Aubi\`{e}re, France, \url{jan.bok@uca.fr}
\and
Department of Applied Mathematics, Faculty of Mathematics and Physics, Charles University, Prague, Czech Republic, \url{{fiala,jedlickova,honza,mikina}@kam.mff.cuni.cz}
}

\begin{document}

\maketitle

\begin{abstract}
The notion of graph covers is a discretization of covering spaces introduced
and deeply studied in topology. In discrete mathematics and theoretical
computer science, they have attained a lot of attention from both the
structural and complexity perspectives. Nonetheless, disconnected graphs were
usually omitted from the considerations with the explanation that it is
sufficient to understand coverings of the connected components of the target
graph by components of the source one. However, different (but equivalent)
versions of the definition of covers of connected graphs generalize to
non-equivalent definitions for disconnected graphs. The aim of this paper is to
summarize this issue and to compare three different approaches to covers of
disconnected graphs: 1) locally bijective homomorphisms, 2) globally
surjective locally bijective homomorphisms (which we call \emph{surjective covers}), and
3) locally bijective homomorphisms which cover every vertex the same number
of times (which we call \emph{equitable covers}). The standpoint of our comparison is
the complexity of deciding if an input graph covers a fixed target graph. We
show that both surjective and equitable covers satisfy what certainly is a natural and
welcome property: covering a disconnected graph is polynomial-time decidable
if such it is for every connected component of the graph, and it is
NP-complete if it is NP-complete for at least one of its components. We further argue that the third variant, equitable covers, is the most natural one,
namely when considering covers of colored graphs. Moreover, the complexity of
surjective and equitable covers differ from the fixed parameter complexity point of
view. 

In line with the current trends in topological graph theory, as well as its applications in mathematical physics, we consider graphs in a very general sense: our graphs may contain loops, multiple edges and also semi-edges. Moreover, both vertices and edges may be colored, in which case the covering projection must respect the colors. We conclude the paper by a complete characterization of the complexity
of covering 2-vertex colored graphs, and show that poly-time/NP-completeness dichotomy holds true for this case.
We actually aim for a stronger dichotomy. All our polynomial-time algorithms work for arbitrary input graphs, while the NP-completeness theorems hold true even in the case of simple input graphs. 
 
\end{abstract}

\section{Introduction}\label{sec:Intro}

The notion of graph covering is motivated by the notion of covering of topological spaces. It has found numerous applications in graph theory, in construction of highly symmetric graphs of requested further properties (cf.~\cite{k:Biggs74,n:MalnicNS00}), but also in models of local computation (\cite{n:Angluin80,n:ChMZ06,n:ChFHPT13}). The application in computer science led Abello, Fellows, and Stilwell~\cite{n:AFS91} to pose the problem of characterizing those (multi)graphs for which one can decide in polynomial time if they are covered by an input graph. They have pointed out that because of the motivation coming from topology, it is natural to consider graphs with multiple edges and loops allowed. Kratochv\'{\i}l, Proskurowski, and Telle~\cite{n:KPT97a} further showed that in order to fully characterize the complexity of covering simple graphs, it is necessary but also sufficient to characterize the complexity of covering colored mixed multigraphs of minimum degree at least three.  In modern topological graph theory it has now become standard to consider graphs with semi-edges since these occur naturally in algebraic graph reductions (informally, a semi-edge has, in contrast to normal edges and loops, only one endpoint.). Bok et al.\ initiated the study of the computational complexity of covering graphs with semi-edges in~\cite{DBLP:conf/mfcs/Bok0HJK21}.   

In all the literature devoted to the computational aspects of graph covers, only covers of connected graphs have been considered so far. The authors of~\cite{n:KPT94} justify this by claiming in Fact~2.b that ``For a disconnected graph $H$, the $H$-cover problem is polynomially solvable
(NP-complete) if and only if the $H_i$-cover problem is polynomially solvable (NP-complete)
for every (for some) connected component $H_i$ of $H$.'' Though this 
seems to be a plausible and desirable property, a closer look shows that the validity of this statement depends on the exact definition of covers for disconnected graphs.

The purpose of this paper is to give this closer look at covers of disconnected graphs in three points of view: the definition, complexity results, and the role of disconnected subgraphs in colored multigraphs. In Section~\ref{sec:whatis} we first discuss what are the possible definitions of covers of disconnected graphs -- locally bijective homomorphisms are a natural generalization from the algebraic graph theory standpoint, globally surjective locally  bijective homomorphisms (which we call {\em surjective covers}) seem to have been understood by the topological graph theory community as the generalization from the standpoint of topological motivation, and a novel and more restrictive definition of {\em equitable covers}, in which every vertex of the target graph is required to be covered by the same number of vertices of the source one. The goal of the paper is to convince the reader that the most appropriate definition is the last one. In Section~\ref{sec:complexity} we inspect the three possible definitions under the
magnifying glass
of computational complexity. The main result is that the above mentioned Fact~2.b is true for surjective covers, and remains true also for the newly proposed definition of equitable covers of disconnected graphs. The NP-hardness part of the statement is proven for instances when the input graphs are required to be simple. Lastly, in Section~\ref{sec:2vertex} we review the concept of covers of colored graphs and show that in this context the notion of equitable covers is indeed the most natural one. We justify our approach by providing a characterization of polynomial/NP-complete instances of the {\sc $H$-Cover} problem for colored graphs with two vertices. It is worth noting that in Section~\ref{sec:connected} we 
do not only summarize the definitions and results on covers of connected graphs, but 
also introduce a new notion of {\em being stronger}, a relation between  connected graphs that generalizes the covering relation and which we utilize in the NP-hardness reductions in Section~\ref{sec:complexity}. We believe that this notion is interesting on its own and that its further study would deepen the understanding of graph covers. 

\section{Covers of connected graphs}\label{sec:connected}

In this section we formally define what we call {\em graphs}, we review the notion of a covering projection for connected graphs and we introduce a quasi-ordering of connected graphs defined by the existence of their simple covers.

\subsection{Graphs with multiple edges, loops and semi-edges}\label{subsec:graphs}

Recall that we allow graphs to have multiple edges, loops and semi-edges.  
A very elegant description of
this notion of a graph through the concept of {\em darts} is used in more algebraic-based papers on covers.
The following formal definition is inspired by
the one given in~\cite{nedela_mednykh}.

\begin{definition}\label{def:graph} 
A \emph{graph} is a triple $(D,V,\Lambda)$, where $D$ is a set of \emph{darts},
and $V$ and $\Lambda$ are each a partition of $D$ into disjoint sets.
Moreover, all sets in $\Lambda$ have size one or two, while in $V$ we allow any number of empty sets (they correspond to isolated vertices).

\emph{Vertices} are the sets of darts forming the partition $V$.
The set of \emph{links} $\Lambda$ splits into three disjoint sets $\Lambda=E\cup L \cup S$, where $E$ represents the \emph{normal edges}, i.e., those links of $\Lambda$ 
that intersect two distinct vertices from $V$, 
$L$ are the \emph{loops}, i.e., those 2-element sets of $\Lambda$ that are subsets of some set from $V$, 
and $S$ are the \emph{semi-edges}, i.e., the 1-element sets from $\Lambda$.
\end{definition}

The standard terminology that a vertex $v\in V$ is \emph{incident} with a link (edge) $e\in \Lambda$
or that distinct vertices $u$ and $v$ are \emph{adjacent} can be expressed as $v\cap e \ne \emptyset$
and as $\exists e \in \Lambda: u\cap e\ne \emptyset \land v\cap e\ne \emptyset$, respectively.

\begin{figure}
\centering
\includegraphics[width=0.9\textwidth]{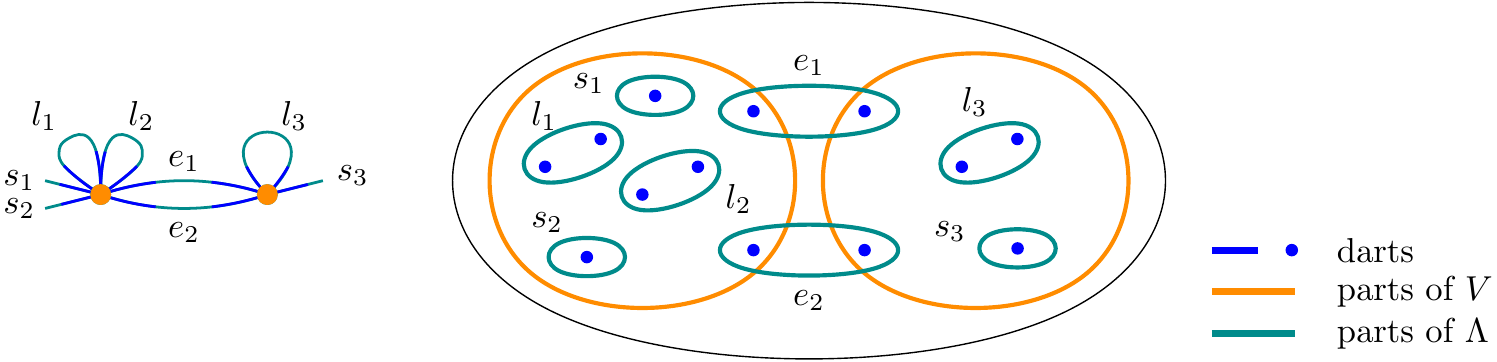}
\caption{An example of a graph drawn in a standard graph-theoretical way (left) and using the alternative dart-based definition (right). The figure appeared first in~\cite{DBLP:conf/mfcs/Bok0HJK21}.}
\label{fig:different-graph-definitions}
\end{figure}

In the standard model a graph is usually defined as an ordered triple $(V,\Lambda,\iota)$, for $\Lambda=E\cup L\cup S$, where $\iota$ is the {\em incidence mapping} $\iota:\Lambda\longrightarrow V\cup{V\choose{2}}$ such that $\iota(e)\in V$ for all $e\in L\cup S$ and $\iota(e)\in {V\choose{2}}$ for all $e\in E$. 
We use both approaches in this paper and employ advantages of each of them in different situations. See an illustrative example in Figure~\ref{fig:different-graph-definitions}. 

To see that both approaches are equivalent we show how the dart representation of a graph can be converted into the incidence one, and vice versa. First, given $G=(D,V,\Lambda)$, we define 
$\iota(e)=\{v: e\cap v \ne \emptyset\}$. For the reverse transformation, given $G=(V,E\cup L\cup S,\iota)$, we define the set of darts as $D=\{(\iota(e),e): e\in S\cup L\} \cup\{ (v,e): v\in\iota(e), e\in E\}$ (with a slight abuse of notation, for every loop $e\in L$, we actually add two copies of $(\iota(e),e)$ into $D$), and then the partition $V$ is given by the equivalence relation $\sim_V$: $(v_1,e_1)\sim_V(v_2,e_2)$ if $v_1=v_2$, and the partition $\Lambda$ by $\sim_\Lambda$: $(v_1,e_1)\sim_\Lambda(v_2,e_2)$ if $e_1=e_2$.

The \emph{degree} of a vertex $v\in V$ is $\deg(v)=|v|$.
The fact that a loop contributes 2 to the degree of its vertex may seem not automatic at first sight, but becomes natural when graph embeddings on surfaces are considered.

The \emph{multiedge} between distinct vertices $u$ and $v$ is the inclusion-wise maximal subset of links that connect $u$ and $v$, i.e. $\{e\in E: e\cap v \ne \emptyset \land v\cap e\ne \emptyset\}$ and the size of this set is the \emph{multiplicity} of the (normal) edge $uv$.
In a similar way we  define the multiplicity of a loop or of a semi-edge.

A graph is \emph{simple} if it has no loops or semi-edges and if every edge has multiplicity one. In this case we use also the standard notation for an edge as $e=uv$ and write $G=(V,E)$.

A graph $H$ is a {\em subgraph} of a graph $G$ if $V(H)\subseteq V(G), E(H)\subseteq E(G), L(H)\subseteq L(G), S(H)\subseteq S(G)$ and $\iota_H(e)=\iota_G(e)$ for every $e\in \Lambda(H)$. The subgraph $H$ is {\em induced} if it is inclusion-wise maximal with respect to the set of links  on the set $V(H)$ of vertices. The subgraph of $G$ induced by a set $W$ of vertices is denoted by $ 
G[W]$. (Note that for the definition of these two notions we have moved to the standard model of graphs, where they are easier to define.)

A \emph{path} in a graph $G$ is a sequence $\ldots,v_i,e_i,v_{i+1},e_{i+1}, \ldots$ of distinct
vertices and links such that for each consecutive triple $v_i,e_i,v_{i+1}$, $\iota(e_i)=\{v_i,v_{i+1}\}$, and for each consecutive triple $e_i,v_{i+1},e_{i+1}$, $\{v_{i+1}\}=\iota(e_i)\cap \iota(e_{i+1})$. Moreover, if the path starts or ends with a link, then this link is a semi-edge; all inner links are normal edges. The path is {\em closed} if it starts and ends with vertices, it is {\em open} if it starts and ends with semi-edges, and it is {\em half-way} in the remaining cases. 

By a \emph{component} of a graph we mean an inclusion-wise maximal induced subgraph such that every two of its vertices are connected by a subgraph isomorphic to a path. 
We say that a graph is \emph{connected} if it has only a single component.

It shall be useful for our purposes to specifically denote one-vertex and two-vertex graphs. Let us denote by $F(b,c)$ the one-vertex graph with $b$ semi-edges and $c$ loops and by $W(k,m,\ell,p,q)$ the two-vertex graph with $k$ semi-edges and $m$ loops at one vertex, $p$ loops and $q$ semi-edges at the other one, and $\ell>0$ multiple edges connecting the two vertices (these edges are referred to as {\em bars}). In other words, $W(k,m,\ell,p,q)$ is obtained from the disjoint union of $F(k,m)$ and $F(q,p)$ by connecting their vertices by $\ell$ parallel edges. Note that the graph in Figure~\ref{fig:different-graph-definitions} is in fact $W(2,2,2,1,1)$. We denote by $G+H$ the disjoint union of (isomorphic copies of) graphs $G$ and $H$, e.g., $W(k,m,0,p,q)=F(k,m)+F(q,p)$.

\subsection{Covers of connected graphs}

Though there is no ambiguity in the definition of graph covers of connected graphs, the standard definition used e.g. in~\cite{n:KPT97a} or~\cite{k:MN98} becomes rather technical especially when semi-edges are allowed. The following simple-to-state yet equivalent definition was introduced in~\cite{DBLP:conf/mfcs/Bok0HJK21}. 

\begin{definition}\label{def:graph-dart-cover}
We say that a graph $G=(D_G,V_G,\Lambda_G)$ \emph{covers} a connected graph $H=(D_H,V_H,\Lambda_H)$  if there exists a surjective mapping 
$f: D_G\to D_H$ such that:
\begin{itemize}
\item For every $u\in V_G$, there is a $u'\in V_H$ such that the restriction of $f$ onto $u$ is a bijection between $u$ and $u'$.
\item For every $e\in \Lambda_G$, there is an $e'\in \Lambda_H$ such that $f(e)=e'$.
\end{itemize}
\end{definition}

We write $G\longrightarrow H$ to express that $G$ covers $H$ when  $H$ is a connected graphs. 

This compact and succinct definition emphasizes the usefulness of the dart definition of graphs
in contrast with the lengthy and technical definition of covers in the standard way which is recalled in the following proposition. Note that it follows straightforwardly from the definition that the mapping of vertices induced by a covering projection is degree-preserving.

\begin{proposition}\label{prop:covering}
A graph $G$ covers a graph $H$ if and only if $G$ allows 
a pair of mappings $f_V:V(G)\longrightarrow V(H)$ and $f_\Lambda:\Lambda(G)\longrightarrow \Lambda(H)$ such that
\begin{enumerate}
\item $f_\Lambda(e)\in L(H)$ for every $e\in L(G)$
 and $f_\Lambda(e)\in S(H)$ for every $e\in S(G)$,
\item $\iota(f_\Lambda(e))=f_V(\iota(e))$ for every $e\in L(G)\cup S(G)$,
\item for every link $e\in \Lambda(G)$ such that $f_\Lambda(e)\in S(H)\cup L(H)$ and $\iota(e)=\{u,v\}$, we have $\iota(f_\Lambda(e))=f_V(u)=f_V(v)$,
\item for every link $e\in \Lambda(G)$ such that $f_\Lambda(e)\in E(H)$ and $\iota(e)=\{u,v\}$ (note that it must be $f_V(u)\neq f_V(v)$), we have $\iota(f_\Lambda(e))=\{f_V(u),f_V(v)\}$,
\item for every loop $e\in L(H)$, $f^{-1}(e)$ is a disjoint union of loops and cycles spanning all vertices $u\in V(G)$ such that $f_V(u)=\iota(e)$,
\item for every semi-edge $e\in S(H)$,  $f^{-1}(e)$ is a disjoint union of edges and semi-edges spanning all vertices $u\in V(G)$ such that $f_V(u)=\iota(e)$, and
\item for every edge $e\in E(H)$,  $f^{-1}(e)$ is a disjoint union of edges (i.e., a matching) spanning all vertices $u\in V(G)$ such that $f_V(u)\in\iota(e)$.\qed
\end{enumerate}
\end{proposition}  

For the convenience of the reader, we add another alternative view on graph covering projections, again formulated in the standard model.

\begin{proposition}\label{prop:covering2}
A graph $G$ covers a graph $H$ if and only if $G$ allows 
a pair of mappings $f_V:V(G)\longrightarrow V(H)$ and $f_\Lambda:\Lambda(G)\longrightarrow \Lambda(H)$ such that
\begin{enumerate}
\item the mappings $f_V$ and $f_{\Lambda}$ are incidence preserving,
\item the preimage $f^{-1}_{\Lambda}(e)$ of a normal edge $e\in E(H)$ such that $\iota(e)=\{u,v\}$ is a matching in $G$ spanning $f^{-1}_V(u)\cup f^{-1}_V(v)$, each edge of the matching being incident with one vertex in   $f^{-1}_V(u)$ and one in  $f^{-1}_V(v)$,
\item  the preimage $f^{-1}_{\Lambda}(e)$ of a loop $e\in L(H)$ such that $\iota(e)=\{u\}$ is a disjoint union of cycles in $G$ spanning $f^{-1}_V(u)$ (both a double edge and a loop are considered to be cycles as well),
\item the preimage $f^{-1}_{\Lambda}(e)$ of a semi-edge $e\in S(H)$ such that $\iota(e)=\{u\}$ is a disjoint union of semi-edges and normal edges in $G$ spanning $f^{-1}_V(u)$. \qed
\end{enumerate}
\end{proposition} 

It follows from Proposition~\ref{prop:covering2}.2. that the preimages of two adjacent vertices have the same size. More generally, if $G$ covers a connected graph $H$ via a covering projection $f$, then $|f^{-1}(u)|=k$ for every $u\in V(H)$, where $k=\frac{|V(G)|}{|V(H)|}$ is an integer. Here the connectedness of $H$ is crucial.  

As the first examples, we include the following observations whose proofs are based on K\"onig-Hall and Petersen theorems on factorization of regular graphs (cf.~\cite{DBLP:conf/mfcs/Bok0HJK21}).

\begin{observation}\label{obs:firstexamples}
\begin{enumerate}
\item For every non-negative integer $k$, a simple graph $G$ covers $F(k,0)$ if and only if $G$ is $k$-regular and  $k$-edge-colorable.
\item For every non-negative integer $k$, a simple graph $G$
covers $F(1,k)$ if and only if $G$ is $(2k+1)$-regular and contains a perfect matching.
\item For every non-negative integer $k$, a simple graph $G$ covers $W(0,0,k,0,0)$ if and only if $G$ is bipartite and $k$-regular. \qed  
 \end{enumerate}
\end{observation} 

The computational problem {\sc $H$-Cover} in whose complexity we are mainly interested in, is defined as follows:

\computationproblem{\sc $H$-Cover}{A graph $G$.}{Does $G$ cover $H$?}

\subsection{A special relation regarding covers}

Graph covering is a transitive relation among connected graphs. Thus when $A\longrightarrow B$ for connected graphs $A$ and $B$, every graph $G$ that covers $A$ also covers $B$.  Surprisingly, the conclusion may hold true also in cases when $A$ does not cover $B$, if we only consider simple graphs $G$. To describe this phenomenon, we introduce the following definition, which will prove useful in several reductions later on.  

\begin{definition}\label{def:stronger}
Given connected graphs $A,B$, we say that $A$ {\em is stronger than} $B$, and write $A\triangleright B$, if every simple graph that covers $A$ also covers $B$.
\end{definition}

The smallest nontrivial example of such a pair of graphs are two one-vertex graphs: $F(2,0)$ with a pair of semi-edges and $F(0,1)$, one vertex with a loop. While $F(0,1)$ is covered by any cycle, only cycles of even length cover $F(2,0)$. So $F(2,0)\triangleright F(0,1)$. More generally, for every $k,p\ge 0 $ and $h>0$, $F(k+2h,p)\triangleright F(k,h+p)$.

\begin{observation}
It follows from the definition, that whenever $A$ is simple, then $(A\triangleright B)$ if and only if  $(A\longrightarrow B)$. \qed
\end{observation}

One might also notice that $\triangleright$ is transitive and thus defines a quasi-order on connected graphs. Many pairs of graphs are left incomparable with respect to this relation, even those covering a common target graph. On the other hand, the equivalence classes of pair-wise comparable graphs may be nontrivial, and the graphs within one class might have different numbers of vertices. For example, $W(0,0,2,0,0)$ and
$F(2,0)$ form an equivalence class of $\triangleright$, as for both of these graphs, the class of simple graphs covering them is exactly the class of even cycles. We believe the relation of $\mbox{being stronger}$ is a concept interesting on its own. In particular, the following question remains open and seems relevant.

\begin{problem}
Do there exist two $\triangleright$-equivalent graphs such that none of them covers the other one?
\end{problem}

So far all examples of $A\triangleright B$ we know are such that either $A\longrightarrow B$ or $A$ contains semi-edges. In the open problem session of GROW 2022, we have formulated this as a conjecture:

\begin{conjecture}(\cite{n:GROW2022-open})
If $A$ has no semi-edges, then  $A \triangleright B$ if and only if $A\longrightarrow B$.
\end{conjecture}

This conjecture has been justified for both  one-vertex cubic graphs $B=F(3,0)$ and $B=F(1,1)$ (and arbitrary $A$) in~\cite{KN2023}, as well as for $A=W(0,0,k,0,0)$ and arbitrary $k$ and $B$. Simple graphs which are witnesses for $A\not\triangleright B$ are called {\em generalized snarks} in there. This is explained by the fact that snarks are 2-connected cubic graphs which are not 3-edge-colorable. It is well known that every 2-connected cubic graph contains
a perfect matching, and thus snarks are witnesses of $F(1,1)\not\triangleright F(3,0)$ (cf. Observation~\ref{obs:firstexamples} and 1. and 2. therein).

\section{What is a cover of a disconnected graph?}\label{sec:whatis}

Throughout this section and the rest of the paper
we assume that we are given two (possibly disconnected) graphs $G$ and $H$ and we are interested in determining whether $G$ covers $H$. In particular in this section we discus what it means that $G$ covers $H$. We assume that $G$ has $p$ components of connectivity, $G_1, G_2, \ldots, G_p$, and $H$ has $q$ components, $H_1, H_2,\ldots, H_q$. It is reasonable to request that a covering projection must map each component of $G$ onto some component of $H$, and this restricted mapping must be a covering. The questions we are raising are:
\begin{enumerate}
\item Should the covering projection be globally surjective, i.e., must the preimage of every vertex of $H$ be nonempty?
\item Should the preimages of the vertices of $H$ be of the same size? 
\end{enumerate}
Both these questions are the first ones at hand when trying to generalize graph covers to disconnected graphs, since the answer is ``yes" in the  case of connected graphs. (and it is customary to call a projection that covers every vertex $k$ times a {\em $k$-fold cover}). 

\begin{definition}
Let $G$ and $H$ be graphs and let us have a mapping $f \colon G\longrightarrow H$.
\begin{itemize}
  \item We say that $f$ is a {\em locally bijective homomorphism} of $G$ to $H$ if  for each component $G_i$ of $G$, the restricted mapping $f|_{G_i}:G_i \longrightarrow H$ is a covering projection of $G_i$ onto some component of $H$. We write $G\longrightarrow_{lb}H$ if such a mapping exists.
  \item We say that $f$ is a {\em surjective covering projection} of $G$ to $H$ if  for each component $G_i$ of $G$, the restricted mapping $f|_{G_i}:G_i \longrightarrow H$ is a covering projection of $G_i$ onto some component of $H$, and $f$ is surjective. We write $G\longrightarrow_{sur}H$ if such a mapping exists.
  \item We say that $f$ is an {\em equitable covering projection} of $G$ to $H$ if  for each component $G_i$ of $G$, the restricted mapping $f|_{G_i}:G_i \longrightarrow H$ is a covering projection of $G_i$ onto some component of $H$, and for every two vertices $u,v\in V(H)$, $|f^{-1}(u)|=|f^{-1}(v)|$. We write $G\longrightarrow_{equit} H$ if such a mapping exists.
\end{itemize}
\end{definition}

A useful tool both for describing and discussing the variants, as well as for algorithmic considerations, is introduced in the following definition.

\begin{definition}\label{def:covpattern}
Given graphs $G$ and $H$ with components of connectivity $G_1, G_2,$  $\ldots, G_p$, and $H_1, H_2,\ldots, H_q$, respectively, the {\em covering pattern} of the pair $G,H$ is the weighted bipartite graph
$
\mathrm{Cov}(G,H)=(\{g_1,g_2,\ldots,g_p,h_1,h_2,\ldots,h_q\},\{g_ih_j:G_i\longrightarrow H_j\})
$
with edge weights 
$
r_{ij}=r(g_ih_j)=\frac{|V(G_i)|}{|V(H_j)|}.
$
\end{definition}

The following observation follows directly from the definitions, but will be useful in the computational complexity considerations.

\sloppy
\begin{observation}\label{obs:different-covers-pattern}
Let $G$ and $H$ be graphs. Then the following statements are true.
\begin{itemize}
  \item We have $G\longrightarrow_{lb}H$ if and only if the degree of every vertex $g_i, i=1,2,\ldots,p$ in $\mathrm{Cov}(G,H)$ is greater than zero.
  \item We have $G\longrightarrow_{sur}H$ if and only if the degree of every vertex $g_i, i=1,2,\ldots,p$ in $\mathrm{Cov}(G,H)$ is greater than zero and $\mathrm{Cov}(G,H)$ has a matching of size $q$.
  \item We have $G\longrightarrow_{equit}H$ if and only if  $\mathrm{Cov}(G,H)$ has a spanning subgraph $\mathrm{Map}(G,H)$ such that every vertex $g_i,i=1,2,\ldots,p$ has degree 1 in $\mathrm{Map}(G,H)$ and for every vertex $h_j$ of $\mathrm{Cov}(G,H)$,
$$\sum_{i:g_ih_j\in E(\mathrm{Map}(G,H))}r_{ij}=k,$$ 
$\mathrm{\ where \ } k=\frac{|V(G)|}{|V(H)|}.$
\end{itemize}
\end{observation}
\fussy

\section{Complexity results}\label{sec:complexity}

We feel the world will be on the right track if {\sc $H$-Cover} is polynomial-time solvable whenever {\sc $H_i$-Cover} is polynomial-time solvable for every component $H_i$ of $H$, while {\sc $H$-Cover} is NP-complete whenever {\sc $H_i$-Cover} is NP-complete for some component $H_i$ of $H$. 
To strengthen the results, we allow arbitrary input graphs (i.e., with multiple edges, loops and/or semi-edges) when considering polynomial time algorithms, while we restrict the inputs to simple graphs when we aim at NP-hardness results. This is in line with the Strong Dichotomy Conjecture, stated in~\cite{iwoca}. In some cases we are able to prove results also from the Fixed Parameter Tractability standpoint. 
In those cases we consider both the source and the target graphs to be a part of the input, and the parameter is typically the maximum size of a component of the target one.   

The following lemma is simple, but useful. Note that though we are mostly interested in the time complexity of deciding $G\longrightarrow H$ for a fixed graph $H$ and input graph $G$, this lemma assumes both the source and the target graphs to be part of the input. The size of the input is measured by the number of edges plus the number of vertices of the input graphs.

\begin{lemma}\label{lem:CovPatt}
Let $\varphi(A,B)$ be the best running time of an algorithm deciding if $A\longrightarrow B$ for connected graphs $A$ and $B$, and let $\varphi(n,B)$ be the worst case of $\varphi(A,B)$ over all connected graphs $A$ of size $n$. Then for given input graphs $G$ and $H$ with components of connectivity $G_1, G_2, \ldots, G_p$, and $H_1, H_2,\ldots, H_q$, respectively, the covering pattern $\mathrm{Cov}(G,H)$ can be constructed in time
$$O(pq\cdot\max_{j=1}^q\varphi(n,H_j))=O(n^2\cdot\max_{j=1}^q\varphi(n,H_j)),$$ where $n$ is the input size, i.e., the sum of the numbers of edges and vertices of $G$ and $H$. \qed
\end{lemma}

\begin{corollary}\label{cor:CovPatt1}
Constructing the covering pattern of input graphs $G$ and $H$ is in the complexity class XP when parameterized by the maximum size of a component of the target  graph $H$, provided the {\sc $H_j$-Cover} problem is polynomial-time solvable for every component $H_j$ of~$H$.
\end{corollary}

\begin{nestedproof}
Suppose the size of every component of $H$ is bounded by $M$. Let ${\cal P}_M$ be the class of all connected graphs $B$ of size at most $M$ such that {\sc $B$-Cover} is decidable in polynomial time. By the assumption, every component $H_j$ of $H$ belongs to ${\cal P}_M$. The class ${\cal P}_M$ is finite (its size depends on $M$), and so there are well defined positive integers $K_M, t_M$ such that $\varphi(n,B)\le K_M\cdot n^{t_M}$ for every $B\in {\cal P}_M$.  
Hence $\varphi(n,H_j)\le K_M\cdot n^{t_M}$ for every $j=1,2,\ldots,q$, and $\mathrm{Cov}(G,H)$ can be constructed in time $O(n^2\cdot n^{t_M})$ by the preceding lemma. \qed
\end{nestedproof}

\begin{corollary}\label{cor:CovPatt2}
The covering pattern of input graphs $G$ and $H$ can be constructed in polynomial time provided all components of $H$ have bounded size and the {\sc $H_j$-Cover} problem is solvable in polynomial time for every component $H_j$ of $H$.
\qed
\end{corollary}

In the following subsections, we discuss and compare the computational complexity of deciding the existence of locally bijective homomorphisms, surjective covers, and equitable covers. The corresponding decision problems  are denoted by {\sc LBHom}, {\sc SurjectiveCover}, and {\sc EquitableCover}. If the target graph is fixed to be $H$, we write {\sc $H$-LBHom}, {\sc $H$-SurjectiveCover}, and {\sc $H$-EquitableCover}, respectively.

\subsection{Locally bijective homomorphisms}

The notion of locally bijective homomorphisms is seemingly the most
straightforward
generalization of the fact that in a graph covering projection to a connected graph ``the closed neighborhood of every vertex of the source graph is mapped bijectively to the closed neighborhood of its image''. However, we show in this subsection that it does not behave as we would like to see it from the computational complexity perspective.
Proposition~\ref{prop:lbhomtriangle} shows
that there are infinitely many graphs $H$ with only two components each such that {\sc $H$-LBHom} is polynomial-time solvable, while {\sc $H_i$-LBHom} is NP-complete for one component $H_i$ of $H$. The polynomial part of the desired properties is, however, fulfilled, even in some cases when both graphs are part of the input: 

\begin{theorem}
If {\sc $H_i$-Cover} is polynomial-time solvable for every component $H_i$ of $H$, then
\begin{enumerate}[label=\roman*)]
  \item the {\sc $H$-LBHom} problem is polynomial-time solvable,
  \item the {\sc LBHom} problem is in XP when parameterized by the maximum size of a component of the target graph $H$, 
  \item the {\sc LBHom} problem is solvable in polynomial time, provided the components of $H$ have bounded sizes.
\end{enumerate}
\end{theorem}

\begin{nestedproof}
i) If $H$ is fixed, it has by itself bounded size, and thus the covering pattern $\mathrm{Cov}(G,H)$ can be constructed in polynomial time by Corollary~\ref{cor:CovPatt2}. As noted in Observation~\ref{obs:different-covers-pattern}, $G\longrightarrow_{lb}H$ if and only if $\mbox{deg}_{\mathrm{Cov}(G,H)}g_i\ge 1$ for all $i=1,2,\ldots,p$, which certainly can be checked in polynomial time, once $\mathrm{Cov}(G,H)$ has been constructed.

ii) Deciding if $G$ allows a locally bijective homomorphism into $H$ is not harder than constructing the covering pattern $\mathrm{Cov}(G,H)$, and this task is in XP when parameterized by the maximum size of a component of the target graph $H$, as shown in Corollary~\ref{cor:CovPatt1}.

iii) Follows straightforwardly from ii). \qed
\end{nestedproof}

However, it is not true that {\sc $H$-LBHom} is NP-complete whenever {\sc $H_j$-Cover} is NP-complete for some component $H_j$ of $H$. Infinitely many examples can be constructed by means of the following proposition. These examples provide another argument for our opinion that the notion of locally bijective homomorphism is not the right generalization of graph covering to covers of disconnected graphs.

\begin{proposition}\label{prop:polywhileNPforonecomponent}
Let $H_1\longrightarrow H_2$ for connected components $H_1, H_2$ of $H=H_1+H_2$, and suppose that {\sc $H_2$-Cover} is polynomial-time solvable. Then {\sc $H$-LBHom} is polynomial-time solvable regardless the complexity of {\sc $H_1$-Cover}.
\end{proposition}

\begin{nestedproof}
Under the assumption $H_1\longrightarrow H_2$, any input graph $G$ allows a locally bijective homomorphism to $H$ if and only if each of its components covers $H_2$. On one hand, if each component of $G$ allows a locally bijective homomorphism to $H_2$, the union of these mappings is a locally bijective homomorphism of $G$ into $H$. On the other hand, if $G$ allows a locally bijective homomorphism into $H$, each component of $G$ covers $H_1$ or $H_2$. However, every component that covers $H_1$ also covers $H_2$. \qed
\end{nestedproof}

There are many examples of pairs of connected graphs $H_1, H_2$ such that $H_1$ covers $H_2$, {\sc $H_1$-Cover} is NP-complete and {\sc $H_2$-Cover} is solvable in polynomial time. In a certain sense it is more interesting that a similar phenomenon as in Proposition~\ref{prop:polywhileNPforonecomponent} may occur even when $H_1$ and $H_2$ are incomparable by covering. 

\begin{figure}[t]
\centering
\includegraphics[scale=1.1]{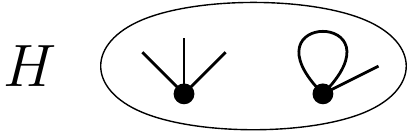}
\caption{The graph $H=H_1+H_2$, where $H_1=F(3,0)$ and $H_2=F(1,1)$.}
\label{fig:Hexample}
\end{figure}

\begin{example}
For graphs $H_1=F(3,0)$ and $H_2=F(1,1)$, neither $H_1$ covers $H_2$ nor $H_2$ covers $H_1$, yet for their disjoint union $H=H_1+H_2$, {\sc $H$-LBHom} is polynomial-time solvable for simple input graphs, while {\sc $H_1$-Cover} is NP-complete for simple input graphs. (The graph $H$ is depicted in Figure~\ref{fig:Hexample}.)
\end{example}

\begin{nestedproof}
A connected simple graph covers $F(3,0)$ if and only if it is cubic and is 3-edge-colorable, in which case it also covers $F(1,1)$ (edges of any two colors form a disjoint union of cycles, which itself covers the loop of $F(1,1)$). Hence a simple graph allows a locally bijective homomorphism to $F(3,0)$+$F(1,1)$ if and only if each of its components covers $F(1,1)$, which can be decided in polynomial time (a connected graph covers $F(1,1)$ if and only if it is cubic and contains a perfect matching). \qed
\end{nestedproof}

This example is a concrete instance of a more general pattern, which in fact has been the reason for introducing the relation $\triangleright$ in Definition~\ref{def:stronger}.

\begin{proposition}\label{prop:lbhomtriangle}
Let $H=H_1+H_2$ for connected graphs $H_1$ and $H_2$ such that $H_1\triangleright H_2$. Then {\sc $H$-LBHom} for simple input graphs
is polynomially reducible to
{\sc $H_2$-LBHom} for simple input graphs. In particular, if {\sc $H_2$-LBHom} is polynomial-time decidable, then so is {\sc $H$-LBHom} as well.
\end{proposition}

\begin{nestedproof}
Every component of the input graph, which is assumed to be simple,  allows a locally bijective homomorphism into $H$ if and only if it covers $H_2$, by the assumption $H_1\triangleright H_2$. Hence for a simple graph $G$, we have $G\longrightarrow_{lb} H$ if and only if $G\longrightarrow_{lb} H_2$. \qed
\end{nestedproof}
 
\subsection{Surjective covers}

The notion of surjective covers is favored by topologists since it captures the fact that every vertex (point) of the target graph (space) is covered [Nedela, private communication 2020]. We are happy to report that this notion behaves as we 
would like to see from the point of view of computational complexity.

\begin{theorem}
If {\sc $H_i$-Cover} is polynomial-time solvable for every connected component $H_i$ of $H$, then
\begin{enumerate}
  \item [(i)] the {\sc $H$-SurjectiveCover} problem is polynomial-time solvable,
  \item [(ii)] the {\sc SurjectiveCover} problem is in XP when parameterized by the maximum size of a component of the target graph $H$, and
  \item [(iii)] the {\sc SurjectiveCover} problem is solvable in polynomial time if the components of $H$ have bounded sizes.
\end{enumerate}
\end{theorem}

\begin{nestedproof}\ 
\begin{enumerate}
  \item [(i)] If $H$ is fixed, the covering pattern $\mathrm{Cov}(G,H)$ can be constructed in polynomial time by Corollary~\ref{cor:CovPatt2}. As noted in Observation~\ref{obs:different-covers-pattern}, $G\longrightarrow_{sur}H$ if and only if $\mbox{deg}_{\mathrm{Cov}(G,H)}g_i\ge 1$ for all $i=1,2,\ldots,p$ and $\mathrm{Cov}(G,H)$ has a matching of size $q$, which can be checked in polynomial time, once $\mathrm{Cov}(G,H)$ has been constructed (e.g., by network flow algorithms).
  \item [(ii)] Again we construct the covering pattern $\mathrm{Cov}(G,H)$, which task is in XP when parameterized by the maximum size of a component of the target graph $H$, as shown in Corollary~\ref{cor:CovPatt1}. Checking the degrees of $\mathrm{Cov}(G,H)$ as well as checking if $\mathrm{Cov}(G,H)$ has a matching of size $q$ can be done in time polynomial in $p+q$ and hence also in the size of the input.
  \item [(iii)] Follows straightforwardly from (ii). \qedhere
\end{enumerate} \qed
\end{nestedproof}

For surjective covers, the NP-hardness of the problem of deciding if there is a covering of one component of $H$ propagates to NP-hardness of deciding if there is a surjective covering of entire $H$, even when our attention is restricted to simple input graphs. 

\begin{theorem}\label{thm:WC-NPc}
The {\sc $H$-SurjectiveCover} problem is NP-complete for simple input graphs if {\sc $H_i$-Cover} is NP-complete for simple input graphs for at least one connected component $H_i$ of $H$.
\end{theorem}

\begin{nestedproof}
Without loss of generality suppose that {\sc $H_1$-Cover} is NP-complete for simple input graphs. Let $G_1$ be a simple connected graph for which $G_1\longrightarrow H_1$ is to be tested. We show that there exists a polynomial-time reduction from {\sc $H_1$-Cover} to {\sc $H$-SurjectiveCover}. 
For every $j=2,\ldots,q$, fix a simple connected graph $G_j$ that covers $H_j$ such that $G_j\longrightarrow H_1$ if and only if $H_j\triangleright H_1$ (in other words, $G_j$ is a witness which does not cover $H_1$ when $H_j$ is not stronger than $H_1$). Note that the size of each $G_j$, $j=2,\ldots,q$, is a constant which does not depend on the size of the input graph $G_1$. 
Note also, that since $H$ is a fixed graph, we do not check algorithmically whether $H_j \triangleright H_1$ when picking $G_j$. We are only proving the existence of a reduction, and for this we may assume the relation   $H_j \triangleright H_1$ to be given by a table.

Let $G$ be the disjoint union of $G_j$, $j=1,\ldots,q$.
We claim that $G\longrightarrow_{sur} H$ if and only if $G_1\longrightarrow H_1$. The ``if'' part is clear. We map $G_j$ onto $H_j$ for every $j=1,2,\ldots,q$ by the covering projections that are assumed to exist. Their union is a surjective covering projection of $G$ to $H$.

For the ``only if'' direction, suppose that $f\colon V(G)\longrightarrow V(H)$ is a surjective covering projection. Since $f$ must be globally surjective and $G$ and $H$ have the same number of components, namely $q$, different components of $G$ are mapped onto different components of $H$ by $f$. Define $\widetilde{f}$ by setting $\widetilde{f}(i)=j$ if and only if $f$ maps $G_i$ onto $H_{j}$. Then  $\widetilde{f}$ is a permutation of $\{1,2,\ldots,q\}$. Consider the cycle containing 1. Let it be $(i_1=1, i_2, i_3, \ldots, i_t)$, which means that $G_{i_j}\longrightarrow H_{i_{j+1}}$ for $j=1,2,\ldots,t-1$, and $G_{i_t}\longrightarrow H_{i_{1}}$. By reverse induction on $j$, from $j=t$ down to $j=2$, we prove that $H_{i_j}\triangleright H_1$. Indeed, for $j=t$, $G_{i_t}\longrightarrow H_1$ means that $H_{i_t}$ is stronger than $H_1$, since we would have set $G_{i_t}$ as a witness that does not cover $H_1$ if it were not. For the inductive step, assume that $H_{i_{j+1}}\triangleright H_1$ and consider $G_{i_j}$. Now $G_{i_j}$ covers $H_{i_{j+1}}$ since $\widetilde{f}(i_j)=i_{j+1}$. Because $G_{i_j}$ is a simple graph and $H_{i_{j+1}}$ is stronger than $H_1$, this implies that $G_{i_j}\longrightarrow H_1$. But then $H_{i_j}$ must itself be stronger than $H_1$, otherwise we would have set $G_{i_j}$ as a witness that does not cover $H_1$. The inductive proof concludes; we proved that $H_{i_2}\triangleright H_1$, and hence $G_1\longrightarrow H_1$ follows from the fact that the simple graph $G_1$ covers $H_{i_2}$. \qed
\end{nestedproof}

\subsection{Equitable covers}

As already announced, we wish to argue that equitable covers form the right generalization of covers of connected graphs to covers of disconnected ones. Not only they capture the crucial properties of covers of connected graphs, but they also behave nicely from the computational complexity point of view.

\begin{theorem}\label{thm:kfoldpoly}
The {\sc $H$-EquitableCover} problem is polynomial-time solvable if {\sc $H_i$-Cover} is polynomial-time solvable for every component $H_i$ of $H$.
\end{theorem}

\begin{nestedproof}
First construct the covering pattern $\mathrm{Cov}(G,H)$. Since $H$ is a fixed graph, this can be done in time polynomial in the size of the input, i.e., $G$, as it follows from Corollary~\ref{cor:CovPatt2}. 

Using dynamic programming, fill in a table $M(s,k_1,k_2,\ldots,k_q)$, $s=0,1,\ldots,p$, $k_j=0,1,\ldots,k=\frac{|V(G)|}{|V(H)|}$ for $j=1,2,\ldots,q$,  with values {\sf true} and {\sf false}. Its meaning is that $M(s,k_1,k_2,\ldots,k_q)=\mbox{\sf true}$ if and only if $G_1\cup G_2\cup\ldots\cup G_s$ allows a locally bijective homomorphism $f$ to $H$ such that for every $j$ and every $u\in V(H_j)$, $|f^{-1}(u)|=k_j$. The table is initialized by setting
$$
M(0,k_1,\ldots,k_q) = \left\lbrace
\begin{array}{ll}
\mbox{{\sf true}, } & \mbox{ if } k_1=k_2=\ldots=k_q=0\\
\mbox{{\sf false}, } & \mbox{ otherwise.}
\end{array} 
\right.
$$
In the inductive step assume that all values for some $s$ are filled in correctly, and move on to $s+1$. For every edge $g_{s+1}h_j$ of $\mathrm{Cov}(G,H)$ and every $q$-tuple $k_1,k_2,\ldots,k_q$ such that $M(s,k_1,k_2,\ldots,k_q)=\mbox{\sf true}$, set $M(s+1,k_1,k_2,\ldots,k_j+r_{s+1,j},\ldots,k_q)=\mbox{\sf true}$, provided $k_j+r_{s+1,j}\le k$. Clearly, the loop invariant is fulfilled, and hence $G$ is a $k$-fold (equitable) cover of $H$ if and only if $M(p,k,k,\ldots,k)$ is evaluated {\sf true}.

The table $M$ has $(p+1)\cdot (k+1)^q=O(n^{q+1})$ entries and the inductive step changes $O((k+1)^q\cdot q)$ values. So processing the table can be performed in $O((k+1)^q(1+pq))=O(n^{q+1})$ steps. \qed
\end{nestedproof}

We will show in Theorem~\ref{thm:Whard} that $q$ in the exponent cannot be avoided if both $G$ and $H$ are part of the input. For this situation, we provide a simpler result.

\begin{corollary}
The {\sc EquitableCover} problem is in XP when parameterized by the number $q$ of connected components of $H$ plus the maximum size of a component of the target graph $H$, provided {\sc $H_i$-Cover} is polynomial-time solvable for every component $H_i$ of $H$.
\end{corollary}

\begin{nestedproof}
The algorithm as described in Theorem~\ref{thm:kfoldpoly} is in XP when parameterized by the number $q$ of components of $H$ (needed for processing the table $M$) plus the maximum size of a component of $H$ (needed for computing the covering pattern). \qed
\end{nestedproof}

\begin{theorem}\label{thm:Whard}
The {\sc EquitableCover} problem in W[1]-hard when parameterized by the number of connected components of $H$, even if the sizes of components of the target graph $H$ are bounded and each {\sc $H_i$-Cover} is polynomial-time solvable for every component $H_i$ of $H$.
\end{theorem}

\begin{nestedproof}
We reduce from {\sc Bin Packing in Unary} parameterized by the number of bins. Given $p$ non-negative integers $x_1,x_2,\ldots,x_p$, the task is to partition this set into $q$ disjoint subsets $S_1,S_2,\ldots,S_q\subset \{1,2,\ldots,p\}$ so that for each $j=1,2,\ldots,q$, the sum $\sum_{i\in S_j}x_i$ of the numbers in each set equals $\frac{\sum_{i=1}^px_i}{q}$. Deciding if this is possible is a W[1]-hard problem when parameterized by the number $q$ of the bins, even if the numbers $x_i$ are encoded in unary~\cite{Jansen2013}. 

Given $p,q$ and the numbers $x_i,i=1,2,\ldots,p$, we set the target graph $H$ to be the disjoint union of $q$ one-vertex graphs, each having one loop incident with its vertex (and no other links). For each $i=1,2,\ldots,p$, $G_i$ will be a cycle of length $x_i$, and $G$ will be the disjoint union of $G_i, i=1,2,\ldots,p$. The components $H_j$ of $H$ are of bounded size (one vertex plus one edge), and for each $j$, {\sc $H_j$-Cover} is solvable in polynomial time, since exactly cycles (of arbitrary lengths) cover $H_j$.  

The covering pattern $\mathrm{Cov}(G,H)$ is thus the complete bipartite graph $K_{p,q}$ with edge weights $r_{ij}=x_i$. Hence $G\longrightarrow_{equit} H$ if and only if the input of the Bin Packing problem is feasible.  
\qed
\end{nestedproof}

The NP-hardness theorem holds true as well:

\begin{theorem}
The {\sc $H$-EquitableCover} problem is NP-complete for simple input graphs if {\sc $H_i$-Cover} is NP-complete for simple input graphs for at least one connected component $H_i$ of $H$.
\end{theorem}

\begin{nestedproof}
We proceed in a similar way as in the proof of Theorem~\ref{thm:WC-NPc}. Suppose without loss of generality that {\sc $H_1$-Cover} is NP-complete. We show that there exists a polynomial-time reduction from {\sc $H_1$-Cover} to {\sc $H$-EquitableCover}. For every $j=2,\ldots,q$, fix a simple connected graph $G_j$ that covers $H_j$ such that $G_j\longrightarrow H_1$ if and only if $H_j\triangleright H_1$ (in other words, $G_j$ is the witness which does not cover $H_1$ when $H_j$ is not stronger than $H_1$). For every $j=2,\ldots,q$, we have integers $k_{j}=\frac{|V(G_j)|}{|V(H_j)|}$ which are constants independent of $G_1$.   

Now suppose we are given a simple graph  $G_1$ whose covering of $H_1$ is to be tested. Compute $k=\frac{|V(G_1)|}{|V(H_1)|}$, which can be done in time polynomial in the size of $G_1$. This $k$ should be an integer, since  otherwise we conclude right away that $G_1$ does not cover $H_1$. Set $K$ to be the least common multiple of $k,k_2,\ldots,k_q$, then $K=\Theta(k)=\Theta(|V(G_1|)$. Define $G$ to be the disjoint union of $\frac{K}{k}$ copies of $G_1$ with $\frac{K}{k_j}$ copies of $G_j$ for all $j=2,\ldots,q$. (Note that the number of connected components of $G$ is 
$p=\frac{K}{k}+\sum\nolimits_{j=2}^q\frac{K}{k_j}$ and the size of $G$ is $\Theta(|V(G_1)|+|\Lambda(G_1)|)$.) 
We claim that $G_1$ covers $H_1$ if and only if $G$ equitably covers $H$, and in that case $G$ is a $K$-fold cover of $H$.

The ``only if'' part is clear. We map each copy of $G_j$ onto $H_j$ for every $j=1,2,\ldots,q$ by the covering projections that are assumed to exist. Their union is a surjective covering projection of $G$ to $H$. To show that this is an equitable covering projection, we do just a little bit of counting. Since $G_j$ is a $k_j$-fold cover of $H_j$ (here and in the sequel, we write $k_1=k$) and we have $\frac{K}{k_j}$ copies of $G_j$ in $G$, the preimage of each vertex of $H$ in this mapping has size $K$.

For the ``if'' part, assume that $f\colon G\longrightarrow H$ is a $K$-fold covering projection. Every connected component of $G$ must map onto one connected component of $H$, but it may happen that different copies of the same $G_j$ map onto different components of $H$. Still, as we argue below, we can again find a sequence of indices $i_1=1, i_2, \ldots, i_t$ such that for every $j=1,2,\ldots,t-1$, some copy of $G_{i_j}$ is mapped to $H_{i_{j+1}}$ by $f$, and some copy of $G_{i_t}$ is mapped onto $H_1$. Then the proof of $G_1\longrightarrow H_1$ proceeds exactly as in the proof of Theorem~\ref{thm:WC-NPc}. 

If some copy of $G_1$ is mapped onto $H_1$, then $t=1$ and $G_1$ covers $H_1$. Suppose this is not the case. Let $S\subseteq \{1,2,\ldots,q\}$ and  let $\cal S$ be an inclusion-wise minimal set of components of $G$ such that:
\begin{enumerate}[label=\it \alph*)]
\item all copies of $G_1$ are in $\cal S$,
\item if a component from $\cal S$ is mapped onto $H_j$ by $f$, then $j\in S$, and
\item if $j\in S$, then all copies of $G_j$ are in $\cal S$.
\end{enumerate}
The sets $S$ and $\cal S$ are uniquely defined by application of the rules a), b), and c). It follows that if $1\in S$, then a sequence $i_1=1,i_2,\ldots, i_t$ exists. If $1\not\in S$, then $f$ restricted to $\cal S$ is a surjective cover of the disjoint union of $H_j,j\in S$. But it cannot be a $K$-fold cover, because the union of the components in $\cal S$ has $\frac{K}{k}|V(G_1)|+\sum\nolimits_{j\in S}\frac{K}{k_j}|V(G_j)|=K\cdot |V(H_1)|+K\sum\nolimits_{j\in S}|V(H_j)|$ vertices, while $\bigcup\nolimits_{j\in S} H_j$ has $\sum\nolimits_{j\in S}|V(H_j)|$ vertices. This concludes the proof. \qed
\end{nestedproof}

\section{Covering colored two-vertex graphs}\label{sec:2vertex}

In this section we introduce
the last generalization and consider coverings of graphs which come with links and vertices equipped with additional information, which we simply refer to as a color. The requirement is that the covering projection respects the colors, both on the vertices and on the links. This generalization is not purposeless as it may seem. It is shown in~\cite{n:KPT97a} that to fully characterize the complexity of {\sc $H$-Cover} for simple graphs $H$, it is necessary and suffices to understand the complexity of {\sc $H$-Cover} for colored mixed multigraphs of valency greater than 2. The requirement on the minimum degree of $H$ gives hope that the borderline between the easy and hard instances can be more easily described. We will first describe the concept of covers of colored graphs with semi-edges in detail in Subsection~\ref{subsec:colored}, where we also give our final argument in favor of equitable covers. Then we extend the characterization of the computational complexity of covering colored 2-vertex graphs without semi-edges presented in~\cite{n:KPT97a} to general graphs in Subsection~\ref{subsec:2-vertex}.     

\subsection{Covers of colored graphs}\label{subsec:colored}

In this section we return to the dart model of graphs, as it is more convenient for describing colored mixed graphs (graphs that allow both undirected and directed links).

\ifx\proof\inlineproof
\begin{definition}
We say that a graph $G$ is \emph{colored}, if it is equipped with a function 
$c\colon D \cup V \to \mathbb N$.

A colored graph covers a colored graph $H$ if $G$ covers $H$ via a mapping $f$ and this mapping respects the colors, i.e., $c_G=c_H\circ f$ on $D$ and every $u\in V_G$ satisfies 
$c_G(u)=c_H(f(u))$.
\end{definition}
\else
\begin{definition}
We say that a graph $G$ is \emph{colored}, if it is equipped with a function 
$c:D \cup V \to \mathbb N$. Furthermore, a colored graph covers a colored graph $H$ if $G$ covers $H$ via a mapping $f$ which respects the colors, i.e., $c_G=c_H\circ f$ on $D$ and every $u\in V_G$ satisfies 
$c_G(u)=c_H(f(u))$.
\end{definition}
\fi

Note that one may assume without loss of generality that all vertices are of the same color, since we can add the color of a vertex as a shade to the colors of its darts. However, for the reductions described below, it is convenient to keep the intermediate step of coloring vertices as well.

 The final argument that equitable covers are the most proper generalization to disconnected graphs is given by the following observation. (Note that color-induced subgraphs of a connected graph may be disconnected.)

\begin{observation}
Let a colored  graph $H$ be  connected and let $f:G\longrightarrow H$ be a  covering projection. Then $f|_{G_{i,j}}:G_{i,j}\longrightarrow H_{i,j}$ is an equitable covering projection for every two (not necessarily distinct) colors $i,j$, where $G_{i,j},H_{i,j}$ denote the subgraphs of $G$ and $H$ induced by the links $e$ such that $c(e)=\{i,j\}$ (note that $c(e)$ is the set of colors of darts that belong to $e$, i.e., $c(e)=\{c(d_1),c(d_2)\}$ if the link $e$ contains two darts $d_1$ and $d_2$, and $c(e)=\{c(d)\}$ if $e$ is a semi-edge containing the dart $d$).\qed 
\end{observation}
Kratochv\'{\i}l et al.~\cite{n:KPT97a} proved that the existence of a covering between two (simple) graphs can be reduced to the existence of a covering between two colored graphs of minimum degree three. Their concept of \emph{colored directed multigraph} is equivalent to our concept of colored graphs (without semi-edges), namely:
\begin{itemize}  
\item The vertex color encoding the collection of trees (without semi-edges) stemming from a vertex is encoded as the vertex color in the exactly same way. 
\item The link color encoding a subgraph isomorphic to colored induced path between two vertices of degree at least three is encoded as the pair of colors of the edge or a loop 
that is used for the replacement of the path. 
\item
When the path coloring is symmetric, we use the same color twice for the darts of the replaced arc which could be viewed as an undirected edge of the construction of~\cite{n:KPT97a}. 
\item On the other hand, when the coloring is not symmetric and the 
replaced arc hence needed to be directed in~\cite{n:KPT97a}, we use a pair of distinct colors on the two darts, which naturally represents the direction. 
\end{itemize}  

When semi-edges are allowed we must take into account one more possibility.
The color used on the two darts representing a symmetrically colored path with an even number of vertices may be used also to represent a half-way path with the identical color pattern ended by a semi-edge. A formal description follows:

By a \emph{pattern} $P$ we mean a finite sequence of positive integers $(p_1,\ldots,p_k)$. 
A~pattern is symmetric if $p_i=p_{k+1-i}$, and the reverse pattern is defined as $\overline{P}=(p_k,\dots,p_1)$.

The pattern of a closed path $u_0,\{d_1,d_2\},u_1,\ldots,\{d_{2k-1},d_{2k}\},u_k$  in a colored graph $G$ is the sequence of colors 
$c(u_0),c(d_1),c(d_2),c(u_1),c(d_3),c(d_4), c(u_2), \ldots,$ $c(d_{2k}),c(u_k)$.
Analogously we define patterns of open and half-way paths.

\begin{figure}[t]
\centering
\includegraphics[width=0.50\textwidth]{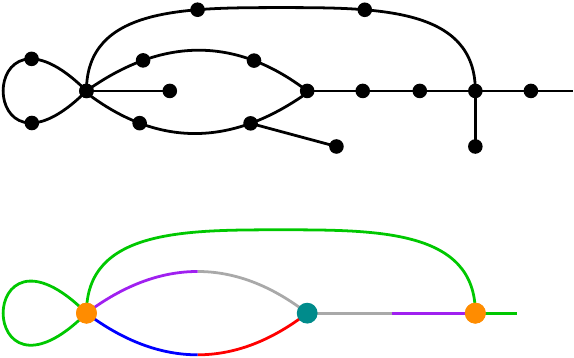}
\caption{Reduction of a graph to a colored graph of minimum degree 3. Distinct colors represent distinct integers.}
\label{fig:cdm}
\end{figure}

Now, a half-way path of pattern $P$ that starts in a vertex of degree 3 and ends by a semi-edge will be replaced by a semi-edge whose color is identical to that used for the two darts forming a normal edge used for the replacement of closed paths whose pattern
is the concatenation  $P\overline{P}$, see Figure~\ref{fig:cdm}.

\subsection{Two-vertex graphs}\label{subsec:2-vertex}

Kratochv\'{\i}l et al.~\cite{n:KPT97a} completely characterized the computational complexity of the {\sc $H$-Cover} problem for colored graphs $H$ with at most two vertices without semi-edges. Their result implies the following:

\begin{proposition}\label{prop:KPT-2-vert} 
Let $H$ be a connected colored graph on at most two vertices without semi-edges.
The {\sc $H$-Cover} problem is polynomial-time solvable if: 
\begin{enumerate}  
\item the graph $H$ contains only one vertex, or
\item $H$ is not regular, or
\item 
\begin{enumerate}  
\item 
for every color $i\in \mathbb N$, the {\sc $H_i$-EquitableCover} problem is solvable in polynomial time, where $H_i$ is the colored subgraph of $H$ induced by the links colored by $i$, and 
\item for every pair of colors $i,j\in \mathbb N$, the {\sc $H_{i,j}$-EquitableCover} problem is solvable in polynomial time, where $H_{i,j}$ is the colored subgraph of $H$ induced by the links $l\in \Lambda$ such that $c(l)=\{i,j\}$.
\end{enumerate}  
\end{enumerate}  
Otherwise, the {\sc $H$-Cover} problem is NP-complete.
\end{proposition}

Informally, the NP-completeness persists if and only if $H$ has two vertices which have the same degree in every color, and the NP-completeness appears on a monochromatic  subgraph (either undirected or directed). Such a subgraph must contain both vertices and be connected. Note explicitly that $H_i$ is a monochromatic undirected subgraph of $H$, while $H_{i,j}$ is often referred to as a monochromatic directed subgraph of $H$. (If $i<j$, we interpret the links of $H_{i,j}$ as colored by the color $(i,j)$, and the links are directed from the dart colored $i$ to the dart colored $j$. In general, $H_{i,j}$ may contain normal edges and loops, but it contains no semi-edges.)  
We extend the characterization from Proposition~\ref{prop:KPT-2-vert} to include semi-edges as well. Because of the results of Section~\ref{sec:complexity}, we restrict our attention to connected target graphs.

\begin{theorem}\label{thm:2-vert}
Let $H$ be a connected colored graph on at most two vertices.
The {\sc $H$-Cover} problem is polynomially solvable if: 
\begin{enumerate}  
\item The graph $H$ contains only one vertex and 
for every $i$, {\sc $H_i$-Cover} is solvable in polynomial time, where $H_i$ is the subgraph of $H$ induced by the loops and semi-edges colored by $i$,
or
\item $H$ is not regular and 
for every $i$ and each vertex $u\in V_H$, the {\sc $H_i^u$-Cover} problem is solvable in polynomial time, where $H_i^u$ is the colored subgraph of $H$ induced by the loops and semi-edges incident with $u$ colored by $i$, 
or
\item $H$ is regular on two vertices and
\begin{enumerate}  
\item  
for every color $i\in \mathbb N$, the {\sc $H_i$-EquitableCover} problem is solvable in polynomial time, where $H_i$ is the colored subgraph of $H$ induced by the links colored by $i$, and 
\item  
for every pair of colors $i,j\in \mathbb N$, the {\sc $H_{i,j}$-EquitableCover} problem is solvable in polynomial time, where $H_{i,j}$ is the subgraph of $H$ induced by the links $l\in \Lambda$ such that $c(l)=\{i,j\}$.
\end{enumerate}  
\end{enumerate}  
Otherwise, the {\sc $H$-EquitableCover} problem is NP-complete.
\end{theorem}

This theorem shows that colored graphs with two vertices exemplify a similar phenomenon as surjective or equitable covers of disconnected graphs -- a polytime/NP-completeness dichotomy applies, and the {\sc $H$-Cover} problem is NP-hard if and only if some monochromatic subgraph induces an NP-hard covering problem. This is in sharp contrast with larger graphs. It has been shown in~\cite{n:KPT97a} that the 3-vertex graph consisting of 2 undirected triangles, each colored by a different color, defines an NP-hard covering problem, while for one color, {\sc $K_3$-Cover} is polynomial-time solvable.

\begin{nestedproof} {\em Of Theorem~\ref{thm:2-vert}.}
We first discuss the polynomial cases:

\begin{enumerate}
\item {\em The graph $H$ has only one vertex and {\sc $H_i$-Cover} is solvable in polynomial time for each $H_i$.} We accept the input if and only if all {\sc $H_{i,j}$-Cover} problems accept the corresponding restricted inputs $G_{i,j}$. Note that for $i\neq j$, $H_{i,j}$ consists of some number of directed loops incident with the vertex of $H$, and such $H_{i,j}$ always defines a polynomial-time solvable covering problem. For undirected monochromatic graphs with one vertex, i.e., $i=j$, {\sc $F(b,c)$-Cover} is polynomial-time solvable if and only if $b\le 1$ or $b=2$ and $c=0$.

In such an admissible case, the overall covering projection $f\colon G\to H$ is the union of all partial covering projections $G_{i,j}\to H_{i,j}$.

\item {\em The graph $H$ has two vertices, it is not regular and {\sc $H_i^u$-Cover} is solvable in polynomial for every $i$ and $u$.} Let the two vertices of $H$ be $v$ and $w$. They can be distinguished: 
\begin{enumerate}
\item by their vertex color, and/or
\item by the number of incident darts of some color $i$.
\end{enumerate}

We perform the same separation on the vertices of $G$ into sets $V_v$ and $V_w$. Namely, $V_v$ contains those vertices of $G$ that have the same color as $v$ and the same number of incident darts of every color as $v$ and analogously for $V_w$. In particular we reject the input if $V_v\cup V_w\ne V(G)$.
We define the vertex mapping $V_G\to V_H$ by mapping the entire $V_v$ to $v$, and $V_w$ to $w$. 

Then, as in the previous case, we check if $G[V_v]$ covers $H^v$, and if $G[V_w]$ covers $H^w$. This can be done in polynomial time according to the assumption.

Lastly we check the covering of edges $e$ incident with both $v$ and $w$.
They can be covered only by edges incident with one vertex in $V_v$ and one in $V_w$.

When $v$ and $w$ are connected by an undirected  multiedge 
of color $i$ (or a directed one of bi-color $(i,j)$) and multiplicity $k$, then a covering may exist if and only if edges of color 
$i$ (or bi-color $(i,j)$, respectively) between $V_v$ and $V_w$ induce a $k$-regular subgraph. 
This necessary condition is also sufficient as every $k$-regular bipartite graph 
can be split into $k$ perfect matchings and these yield the dart mapping of a cover.

\item {\em The graph $H$ has two vertices, is regular, and {\sc $H_{i,j}$-EquitableCover} is solvable in polynomial time for every $i,j$.} Let $V(H)=\{v,w\}$. For every vertex $u\in V(G)$, we introduce a Boolean variable 
$x_u$. Based on the structure of $G$ and $H$ we compose a formula $\varphi$ in CNF with clauses of size two whose  satisfying assignments are in one-to-one correspondence with covering projections from $G$ to $H$. For a covering projection $f\colon G \to H$, 
 $x_u$ is evaluated to {\sf true} if $f(u)=v$, and $x_u$ is evaluated to {\sf false} if $f(u) = w$.
The constitution of $\varphi$ relies on the characterization of polynomially solvable cases 
of the {\sc $H$-Cover} problem for symmetric graphs with two vertices given by Kratochvíl et al.~\cite{n:KPT97a} and Bok et al.~\cite{DBLP:conf/mfcs/Bok0HJK21}. Luckily, all the polynomially solvable cases can be solved via {\sc 2-Sat}, and hence the formula $\varphi$ is simply obtained as the conjunction of subformulas for pairs of (not necessarily distinct) colors $i,j$.
\end{enumerate}

Now we proceed to the NP-complete cases. We will only prove the dichotomy result as stated, i.e., for general inputs, not to overwhelm the reader with technical details of proving NP-hardness for simple input graphs. This will be subject of a strong dichotomy result for a larger class of target graphs in a forthcoming paper~\cite{BFJKS2023?}. The proof below is based on the description of monochromatic graphs that define NP-hard instances of the covering problem:

The problem {\sc $F(a,b)$-Cover} is NP-complete even for simple input graphs when $a\ge 2$ and $a+b\ge 3$~\cite{DBLP:conf/mfcs/Bok0HJK21},

The problem {\sc $W(k,m,\ell,p,q)$-Cover} is NP-complete even for simple bipartite graphs when $\ell\ge 1$, $k+2m=q+2p>0$ and $k+2m+\ell\ge 3$~\cite{DBLP:conf/mfcs/Bok0HJK21},

The problem {\sc $WD(m,\ell,m)$-Cover} is NP-complete for $\ell\ge 1$, $m>0$ and $m+\ell\ge 3$~\cite{n:KPT97a} (here $WD(m,\ell,m)$ is the directed graph with two vertices, $m$ directed loops incident with each of the vertices, and $\ell$ directed edges in each direction between the two vertices).

\begin{enumerate}
\item {\em The graph $H$ has only one vertex.} Let $H_i$ be the subgraph of $H$ for which the {\sc $H_i$-Cover} problem is NP-complete.
By $H'$ we denote the complement of $H_i$ in $H$. Let $G_i$ be the graph for which 
the covering to $H_i$ is questioned. We create a graph $G$ from $G_i$ and $|V(G_i)|$ copies of $H'$ 
by identifying the vertex of each copy of $H'$ with a vertex of $G_i$. Clearly the size of $G$ is $\Theta(G_i)$. We claim that $G$ covers $H$ if and only if $G_i$ covers $H_i$

When $G_i$ covers $H_i$, then we extend the covering to each copy of $H'$ by the identity mapping on $H'$. On the other hand the restriction of a covering projection $G \longrightarrow H$ to the subgraph $G_i$ is a covering projection to $H_i$.

\item {\em The graph $H$ has two vertices and it is not regular.} Let $H^u_i$ be a monochromatic subgraph induced by one of the vertices of $H$ that defines an NP-complete covering problem. We apply the same approach as in the previous case, we just use $H_i^u$ instead of $H_i$.

\item {\em The graph $H$ has two vertices and it is regular.}
In this case, we will exploit the concept of a (``categorical'') graph \emph{product}: For a colored graph $G$, the product $G\times 2$ has as the dart set the Cartesian product $D(G) \times \{1,2\}$. To simplify our expressions we use $d_1$ for $(d,1)$ and $u_1$ for $u\times \{1\}$ when the use of indices cannot be misinterpreted.
The two darts $d_1,d_2$ have the same color as $d$. 
Every vertex 
$u\in V(G)$ gives rise to two vertices $u_1$ and $u_2$ of the same color as $u$.
Every semi-edge $s=\{d\}\in S$ gives rise to a normal edge $\{d_1,d_2\}$, while every loop or normal edge 
$\{d,d'\}\in L \cup E$ gives rise to two normal edges $\{d_1,d'_2\}$ and $\{d'_1,d_2\}$. Note that $G\times 2$ has no semi-edges nor loops, but it may have multiple normal edges.

Observe that mapping both $d_1,d_2$ onto $d$ for every $d\in D$, we get a covering projection $G\times 2 \longrightarrow G$. The graph $G\times 2$ is also referred to as the {\em canonical double cover} of $G$, since any bipartite graph covers $G$ if and only if it covers $G\times 2$. 

\begin{enumerate}
\item {\em Let $H_i$ be a spanning monochromatic subgraph of $H$ for which {\sc $H_i$-Eq\-ui\-table\-Co\-ver} is NP-complete.} Let $H'$ be the complement of $H_i$ in $H$ and let $G_i$ be the graph for which 
the covering to $H_i$ is questioned. 

\begin{itemize}
\item
For connected $H_i$, we may assume that $G_i$ is connected as well. We take two copies of $G_i$, one copy of $G_i \times 2$ 
and $|V(G_i)|$ copies of $H'\times 2$. If $u$ is a vertex of $G_i$, let $u'$ and $u''$ be the two vertices corresponding to $u$ in the two copies of $G_i$, while $u'''_1$ and $u'''_2$ 
be the vertices that arouse from $u$ in $G_i \times 2$. Analogously let vertices $v_1,v_2,w_1,w_2$ be the vertices obtained from $v$ and $w$ of $H'$ in $H'\times 2$. We form $G$ by choosing for each $u\in V(G_i)$ 
a unique copy of $H'\times 2$ and identifying
the pairs 
$(u',v_1)$, $(u'',v_2)$, $(u'''_1,w_1)$ and $(u'''_2,w_2)$, 
where $v_1,v_2,w_1,w_2$ are taken from the chosen copy of $H'\times 2$,
see Figure~\ref{fig:construction}

\begin{figure}
\centering
\includegraphics[page=1,width=0.9\textwidth]{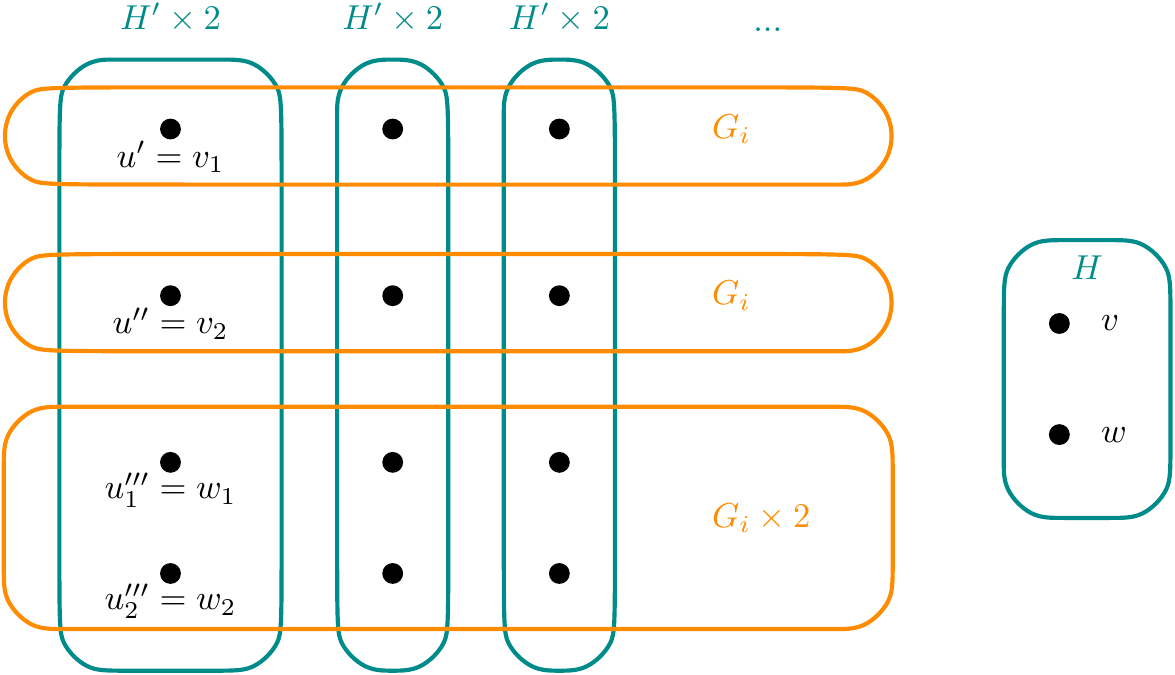}
\caption{Construction of $G$ in Theorem~\ref{thm:2-vert} for connected $H_i$.\label{fig:construction}}
\end{figure}

Observe that if a bipartite graph $\Gamma$ covers the regular 2-vertex graph $H$ via a mapping $f: \Gamma \longrightarrow H$, then the companion mapping (which we call the {\em swap}) $\widetilde{f}$ defined on vertices by $\widetilde{f}(u)=w$ iff $f(u)=v$, also determines a covering projection $\widetilde{f}:\Gamma\longrightarrow H$ (because $\widetilde{f}$ is a degree-obedient vertex mapping, cf.~\cite{DBLP:conf/mfcs/Bok0HJK21}).
  
We claim that $G$ covers $H$ if and only if $G_i$ covers $H_i$.
If $G_i$ covers $H_i$ via a covering projection $f$, then use this $f$ on the 2 copies of $G_i$ and use its swap on $G_i \times 2$, and denote this vertex mapping by $g$. For every $u\in V(G_i)$, we get either $g(u')=g(u'')=v$ and $g(u_1''')=g(u_2''')=w$, or vice versa. Thus on each copy of $H'\times 2$ we have obtained a mapping that can be extended to a covering projection. Their union, together with $g$, is a covering projection from $G$ to $H$.  

On the other hand, the restriction of a covering projection $G\longrightarrow H$ to $G_i$ yields a covering projection $G_i\longrightarrow H_i$.
Since $H_i$ is connected, this is an equitable covering.
(Disconnected $H_i$ would yield only a locally 
bijective homomorphism $G_i\longrightarrow_{lb} H_i$.)

\item
For disconnected $H_i$, we first recall that the {\sc $H_i$-EquitableCover} is polynomially solvable if each component of $H_i$ is incident with at most one semi-edge or the degree of $H_i$ is two.

Let $H_i^+$ be the component of $H_i$ with the maximum number of semi-edges, i.e., at least two, let $v$ be the vertex of $H_i^+$ and let $H_i^-$ be its complement in $H_i$. We reduce from the NP-complete problem {\sc $H_i^+$-Cover}. (Note that {\sc $H_i^-$-Cover} could be polynomially solvable.)

Let us have a
connected graph  $G_i$ as an instance of {\sc $H_i^+$-Cover}. 
We use $G_i$ together with $|V(G_i)|$ copies of $H_i^-$ and 
$|V(G_i)|$ copies of $H'$. 
For each vertex $u\in V(G_i)$, we take a copy of $H'$ and identify its vertex $v$ with $u$, while the other vertex $w$ of this copy of $H'$ is identified with the vertex of $H_i^-$, see Figure~\ref{fig:construction-2}. This concludes the construction of $G$.

\begin{figure}
\centering
\includegraphics[page=2,width=0.9\textwidth]{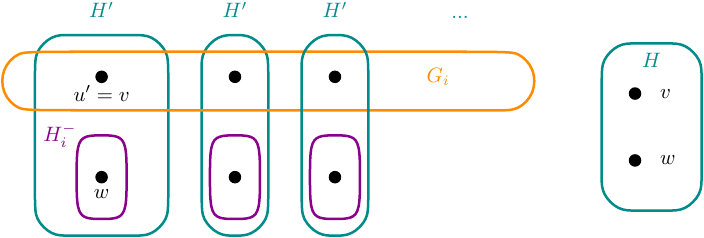}
\caption{The construction of $G$ in Theorem~\ref{thm:2-vert} for disconnected $H_i$.\label{fig:construction-2}}
\end{figure}

Again we claim that $G\longrightarrow H$ if and only if $G_i\longrightarrow H_i^+$. For the ``only if" direction,
observe that a copy of $H_i^-$ may cover $H_i^+$  only if these graphs are isomorphic. Hence we may assume that every copy of $H_i^-$ is mapped  on the subgraph $H_i^-$, and then
the copy of $G_i$ is mapped onto $H_i^+$.

On the other hand, if $G_i\longrightarrow H_i^+$, the desired covering projection $G\longrightarrow H$ is constructed from this mapping on $G_i$, combined with mapping every copy of $H'$ and every copy $H_i^-$ by identity mappings onto $H'$ and   $H_i^-$, respectively.

\end{itemize}
\item {\em Let {\sc $H_{i,j}$-EquitableCover} be NP-complete for some bi-colored subgraph $H_{i,j}$, $i\neq j$.} Note that $H_{i,j}$ is a directed graph, and hence it is connected (since one-vertex directed graphs determine polynomial-time solvable instances of graph covers).
Then we perform the same construction of $G$ 
from two copies of the instance $G_{i,j}$, a copy of $G_{i,j} \times 2$ and 
$|V(G_{i,j})|$ copies of $H'\times 2$ that are merged in the same way as in the connected subcase of 3.a). The arguments are then identical.  \qed
\end{enumerate}
\end{enumerate}

\end{nestedproof}
 
\section{Conclusion}

The main goal of this paper was to point out that the generalization of the notion of graph covers of connected graphs to covers of disconnected ones is not obvious. We have presented three variants, depending on whether the projection should be or does not need to be globally surjective, and if all vertices should  be or do not need to be covered the same number of times. We argue that the most restrictive variant, which we call equitable covers, is the most appropriate one, namely from the point of view of covers of colored graphs. 

We have compared the computational complexity aspects of these variants and show that two of them, surjective and equitable covers, possess the naturally desired property that {\sc $H$-Cover} is polynomially solvable if covering each component of $H$ is polynomially solvable, and NP-complete if covering at least one component of $H$ is NP-complete. 

In the last section we review the extension of graph covers to covers of colored graphs, recall that colors can be encoded by non-coverable patterns in simple graphs, and discuss this issue in detail for the case when semi-edges are allowed. With this new feature we conclude the complete characterization of the computational complexity of covering 2-vertex colored graphs, initiated (and proved for graphs without semi-edges) 24 years ago in~\cite{n:KPT97a}. 

Last but not least, some of the hardness reductions are based on a newly introduced notion of $\triangleright$ order of connected graphs, which expresses inclusions among classes of simple covers of the graphs. We believe that a better understanding of this relation would shed more insight into the concept of graph covers as a whole, and state two open problems about this relation.   

\section*{Acknowledgments}

\begin{itemize}
  \item Jan Bok: Supported by the ANR project GRALMECO (ANR-21-CE48-0004).
  \item Nikola Jedličková: Supported by research grant GAČR 20-15576S of the Czech Science Foundation, by SVV--2020--260578, and GAUK 1580119.
  \item Jiří Fiala and Jan Kratochvíl: Supported by research grant GAČR 20-15576S of the Czech Science Foundation.
  \item Michaela Seifrtová: Supported by research grant GAČR 19-17314J of the Czech Science Foundation.
\end{itemize}

The authors thank Ondra Such\'y for valuable comments, namely for pointing out the reference~\cite{Jansen2013}.

\bibliography{bib/knizky,bib/nakryti,bib/sborniky,0-main}

\end{document}